# Gigapixel microscopy using a flatbed scanner

**Guoan Zheng,[*] Xiaoze Ou, and Changhuei Yang**

*Department of Electrical Engineering, California Institute of Technology, Pasadena, CA 91125, USA*

[*]gazheng@caltech.edu

**Abstract:** Microscopy imaging systems with a very wide field-of-view (FOV) are highly sought in biomedical applications. In this paper, we report a wide FOV microscopy imaging system that uses a low-cost scanner and a closed-circuit-television (CCTV) lens. We show that such an imaging system is capable to capture a 10 mm * 7.5 mm FOV image with 0.77 micron resolution, resulting in 0.54 gigapixels ($10^9$ pixels) across the entire image (26400 pixels * 20400 pixels). The resolution and field curve of the proposed system were characterized by imaging a USAF resolution target and a hole-array target. A 1.6 gigapixel microscopy image (0.54 gigapixel with 3 colors) of a pathology slide was acquired by using such a system for application demonstration.

**OCIS codes:** (170.0110) Imaging systems; (170.4730) Optical pathology; (170.0180) Microscopy

---

## References and links

## 1. Introduction

The conventional microscope architecture can be generally defined as consisting of a microscope objective for light collection from a sample slide, intermediate relay optics and a pair of or a singular eyepiece that projects a magnified image of the sample into our eyes. With the advancement of digital cameras, the eyepiece(s) segment of the microscope has undergone adaptation changes to be replaced with appropriate optics and camera to enable electronic imaging. Over the past decades and with the broad acceptance of infinity correction, the conventional microscope design has achieved extensive standardization across the microscopy industry - objectives and eyepieces from the major microscope makers are largely interchangeable. This standardization helps with cost-effectiveness. However, it has also limited the commercial design space for conventional microscopy - any significant design deviation that exceeds the standardization parameter space would have to contend with its incompatibility with the entrenched microscopy consumer base.

Recently, there has been an increased recognition that bioscience and biomedical microscopy imaging needs are outstripping the capability of the standard microscope. One salient need of modern bioscience and biomedical community is for a microscopy imaging method that is able to electronically acquire a wide field-of-view (FOV) image with high resolution [1]. The standard microscope was originally designed to provide sufficient image details to a human eye or a digital camera sensor chip. As an example, the resolution of a conventional 20X objective lens (0.4 numerical aperture) is about 0.7 µm and the FOV is only about 1 mm in diameter. The resulting space-bandwidth-product (SBP) [2] is about 8 mega-pixels (the number of independent pixels to characterize the captured image). This pixel count has only been recently reached or exceeded by digital camera imager. Interestingly, this SBP shows only slight variation across the range of commercial microscope objectives. Placed in different context, the relative invariance of SBP necessarily ties resolution and FOV together for most commercial objectives - high-resolution imaging necessarily implies a limited FOV.

In the past years, there has been significant progress in the development of system that increases the FOV of the conventional microscope system by incorporating sample slide scanning to acquire image over a large area [3] or by implementing parallel imaging with multiple objectives [4]. In addition, there have also been exciting research efforts into wide FOV imaging system, including ePetri dish [5], digital in-line holography [6], focus-grid scanning illumination [7, 8], off-axis holography microscopy [9, 10]. All these methods try to break the tie between resolution and FOV by abandoning the conventional microscopy design and shifting away from the use of optics schemes that perform optical image magnification.

The underlying assumptions that underpin all of these developments appear to be 1) a higher SBP (order of magnitude or more) with a magnification-based optical scheme is commercially impractical and 2) the associated pixel count for a radically higher SBP would face electronic image acquisition issues for which a viable solution does not yet exist.

In this paper, we demonstrate an optical magnification microscopy solution that challenges these assumptions. The configuration of this imaging system is based on two cost-effective items: a commercial available closed-circuit-television (CCTV) lens system and a low-cost consumer flatbed scanner. We show that, such a system is capable to capture a 0.54 gigapixel microscopy image with a FOV of 10 mm * 7.5 mm and that a 0.77 resolution is achieved across the entire FOV. Remarkably, the CCTV lens system has a SBP of at least 0.5 gigapixel ($10^9$ pixels), which is about 2 orders of magnitude larger than those of conventional microscope objectives.

This paper is structured as follows: we will first present our proof-of-concept setup; then, we will present the automatic focusing scheme of the platform. Next, we will report on the resolution and the field-curve characterization of the platform. Then, we will demonstrate the application of the proposed setup by imaging a blood smear and a pathology slide; finally, we will discuss some limitations as well as future directions for the proposed gigapixel microscopy system.

## 2. The prototype setup of the 0.5 gigapixel microscopy imaging system

Driven by the recent trend of small pixel size of the image sensor (0.7μm pixel size has been reported in Ref. [11]), significant efforts have been put into the design of consumer/industry camera lens to match this diffraction-limited pixel size. In the past years, it has been demonstrated that the SBP of some consumer camera lenses can achieve count on the order of billion pixels, i.e. these camera lenses are capable to capture gigapixel images [12, 13].

In our proposed microscopy imaging system, we redirected this gigapixel imaging effort [12, 13] to microscopy. The main component of the setup was a commercial available high-quality CCTV lens (C30823KP, Pentax, f/1.4, focal length 8mm). Like other consumer/industry camera lenses, the conventional use of this lens is to demagnify the scene onto the image plane, where the CMOS/CCD imager is located. In our setup (Fig. 1), we put our sample at the image plane to replace the CMOS/CCD imager and used the CCTV lens to magnify the sample, i.e. using the lens in the reverse manner. With a magnification factor of ~30, the projected image was too large to be directly imaged with a CMOS/CCD imager. Instead, we modified and employed a consumer flatbed scanner (Canon LiDE 700F) to accomplish image acquisition. We chose this scanner for two reasons: 1) its "LED Indirect Expose (LiDE)" design and 2) the high scanning resolution (2400 dpi is measured, corresponding to a 10 μm pixel size of the scanner). Due to its LiDE design, this scanner actually possesses a linear CCD that covers the complete width of the scanning area. In comparison, other conventional scanners use a combination of mirrors and lenses to accomplish the same functionality, and it will take additional steps to modify these scanners for our application. In our setup, we disabled the LED light source of the Canon LiDE 700F scanner by using a black tape. The relay lens array and the light guide on top of the linear CCD were also removed. Therefore, the linear CCD shown in Fig. 1 was directly exposed to the projected image from the CCTV lens.

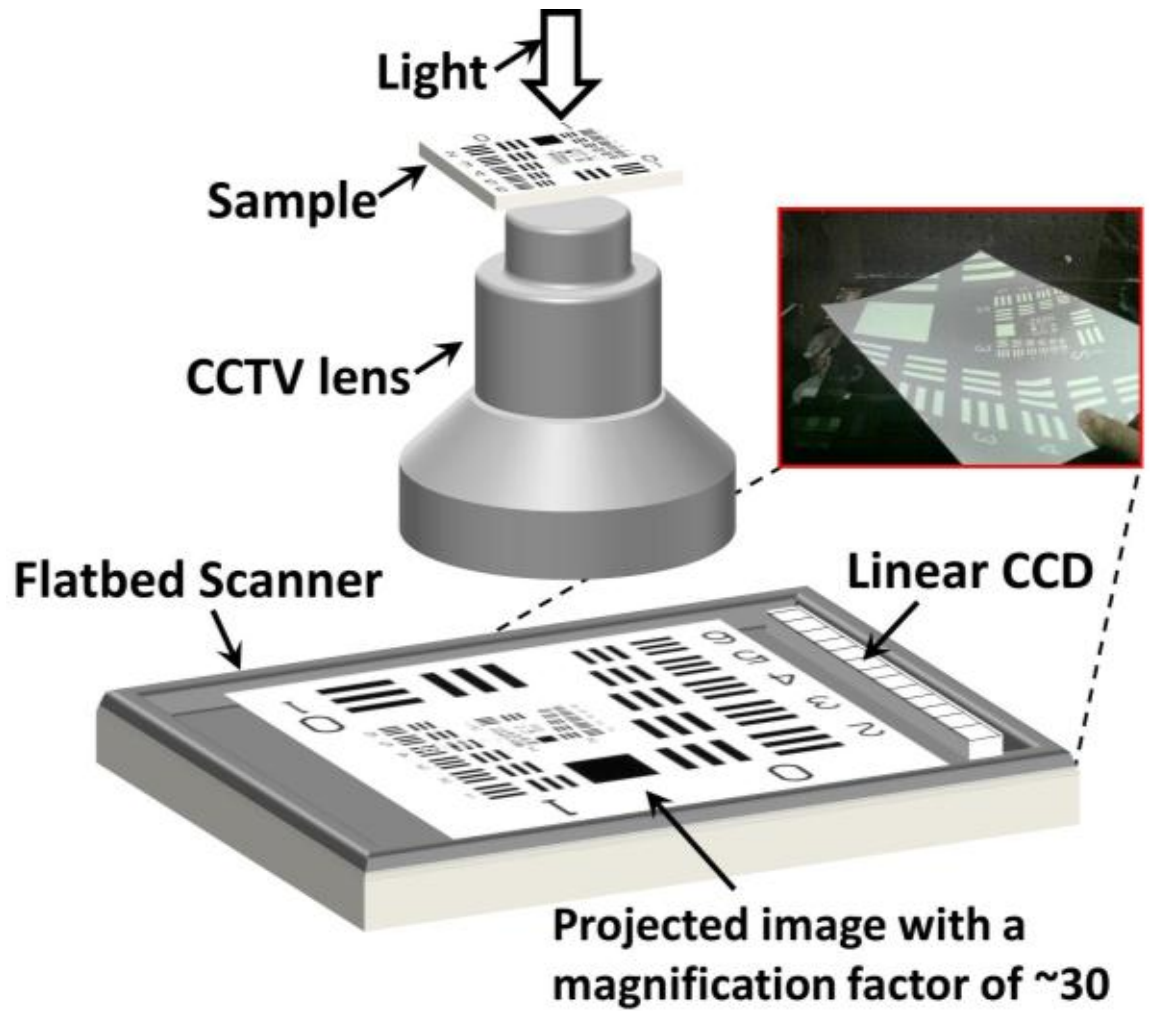

Fig. 1. The setup of the proposed 0.54 gigapixel microscopy (not to scale). A CCTV lens is used to magnify the sample by a factor of 30 and a scanner is used to capture the projected image. The distance between the sample and the lens is about 1 cm. Inset on the top right shows the magnified image of a USAF target on an A4 paper held in front of the scanner.

The scanning resolution was set to 2400 dpi, the maximum resolution of the scanner (corresponds to a 10 μm pixel size); the FOV of the scanner was set to the maximum scanning area (297 mm * 216 mm). The magnification factor was ~30 in our platform, and it corresponded to a FOV of 10 mm * 7.5 mm at the object side. The distance between the sample and the CCTV lens is about 1 cm and the distance between scanner and the lens is about 30 cm. We used a diffuse LED light source from top for illumination. Based on these settings, the capture image contained 26400 pixels * 20400 pixels, and thus, it produced a 0.54 gigapixel microscopy image of the sample. Inset of Fig.1 shows the projected image of a USAF target on a letter size paper held in front of the scanner.

## 3. Automatic focusing scheme

In a conventional microscope setup, the image recording device can give real time update on the object, and therefore, it is easy to adjust the position of the stage to the best focus position. In the proposed platform, the scanner takes a relative long time to acquire the entire image of the object, and thus, it takes a long time to find out the best focus position. To address this issue, we used an automatic focusing scheme with a programmable stage in our setup, as shown in the inset of Fig. 2(a2). This focusing scheme consists of three steps: 1) move the stage with a constant speed (5 μm/s); 2) acquire the image at the same time (only certain part of the image is in-focus along the scanning direction); 3) define an F index to find out the best focus position from the acquire image. In step 3, we define the F index using the following equation

$$F\ index = \sum_{x=1}^{end} \left| 2f(x,y) - f(x-step,y) - f(x+step,y) \right| \quad (1)$$

Such an F index is a measurement of the sharpness of the image.

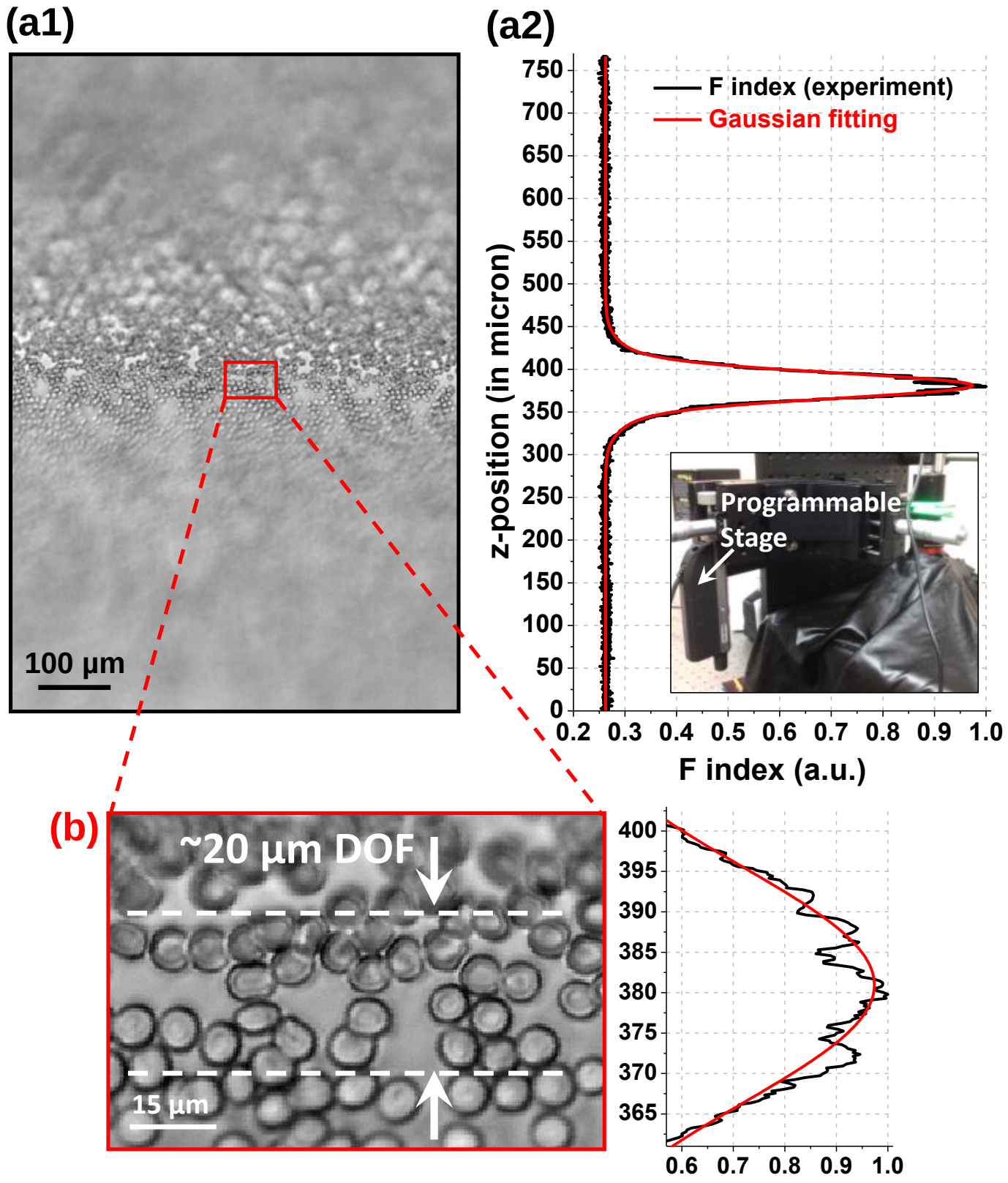

Fig. 2. The automatic focusing scheme of the proposed setup. (a1) The acquired image of a blood smear with the stage moving at a constant speed at z-direction. (b) Based on the motion speed, we can plot the F index with respect to different z positions, and thus, automatically locate the best focused position of the sample. (b) The magnified image of (a), where the depth-of-focus is estimated to be ~20 μm. In this experiment, a diffused green LED with 20 nm spectrum bandwidth is used for illumination.

Fig. 2(a1) is the acquired image of a blood smear following the above steps (only a small portion of the image is acquired for faster scanning). We can see that the sample is out-of-focus at the beginning, then, the stage bring the sample into focus at the middle part of the image, and finally, the sample is out-of-focus again. Based on the motion speed of the programmable stage, we can plot the F index (with '*step*'=2) versus different z-positions, as shown in Fig. 2(a2). The peak of the F index is estimated to be located at z=381 µm. The magnified image and the corresponding F index are shown in Fig. 2(b), where the depth-of-focus (DOF) is estimated to be ~20 µm. The proposed automatic focusing scheme works well with biological samples and pathology slides. The entire focusing process takes about 1~2 minutes for the proposed platform. However, we note that, if the sample is extremely sparse (for example, one small hole on a metal mask), such a scheme will fail and we have to take multiple images at different z positions to find out the best focus position.

## 4. The resolution and the field curve of the platform

We next characterize the resolution and the field-curve of the proposed imaging system. We first imaged a USAF target, as shown in Fig. 3. It is well known that, the aberration of a physical lens will degrade the image resolution at different FOVs of the lens, i.e. the resolution may be different from the center to the edge FOVs. In order to test the resolution at different FOVs, we translated the USAF target across the FOV of the CCTV lens and captured the corresponding images in Fig. 3(a-c). A diffused green LED light source (530 nm wavelength, with ~20 nm spectrum bandwidth) was used in this experiment for illumination.

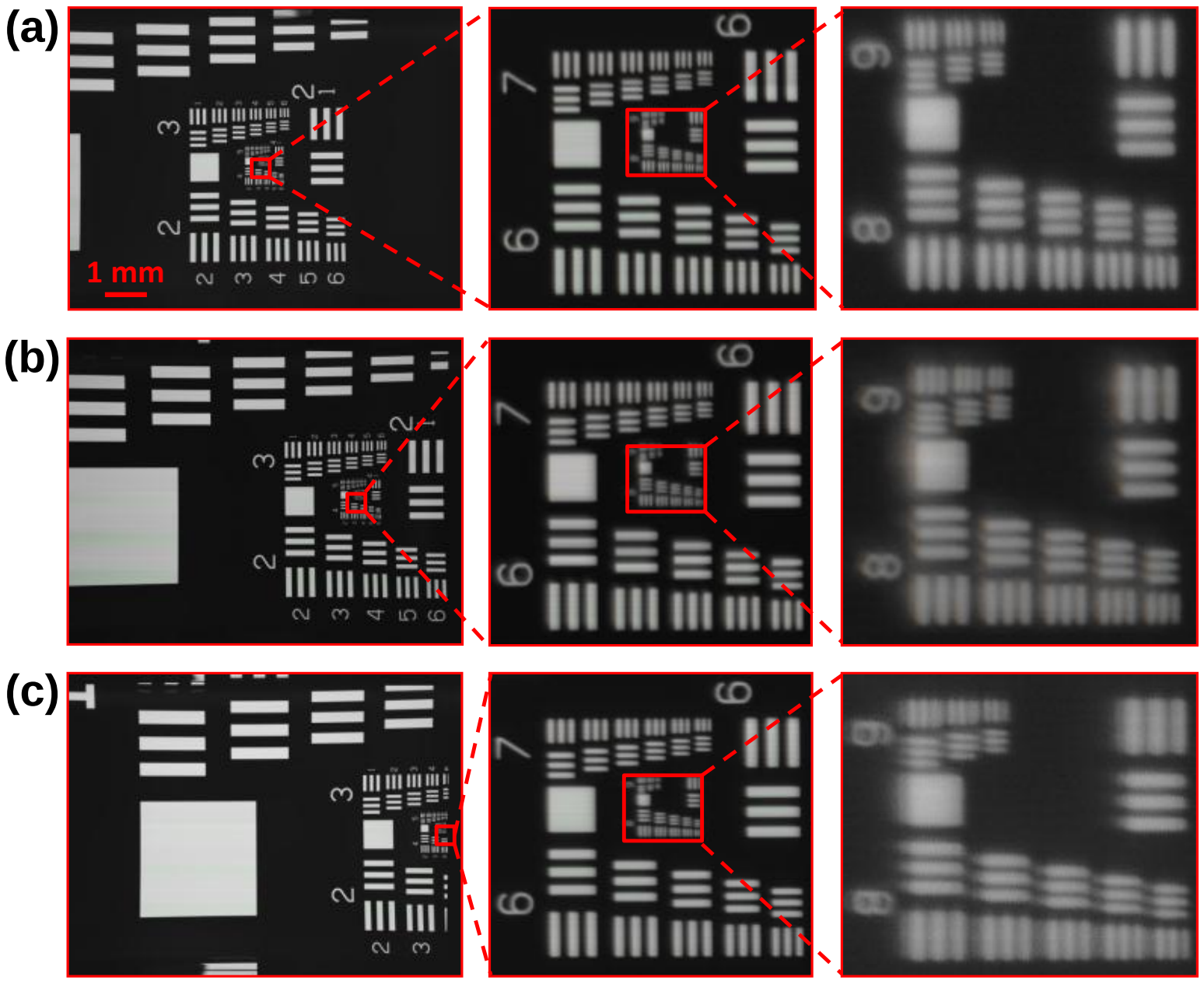

Fig. 3. USAF resolution target acquired by the proposed 0.54 gigapixel microscopy system. The effective FOV is about 10 mm * 7.5 mm, with 26400 pixels * 20400 pixels across the entire image. The imaging performance at the (a) center, (b) 50% away from center and (c) 95% away from center. The line widths of group 9, element 1, 2 and 3 are 0.98 µm, 0.87µm and 0.77 µm respectively.

In the USAF target, the line widths of group 9, element 1, 2 and 3 are 0.98 µm, 0.87µm and 0.77 µm respectively. The image performance at the center of the FOV is shown in Fig. 2(a), where we can clearly see the feature at group 9, element 3 (0.77µm line width). In Fig. 2(b) and (c), we translate the sample to 50% and 95% of the FOV away from center (100% corresponds to 10 mm). In both images, we can still clearly see the fine feature at group 9, element 3 (0.77µm line width). This establishes the resolution of our prototype system is, under the monochromatic green light illumination, 0.77 µm over the entire FOV. We note that, by Nyquist theorem, we need at least two pixels to capture the smallest detail of the image, and thus, the effective pixel size at the object size should be less than 385 nm. Based on section 1, the effective pixel size of our platform is about 330 nm (10 µm divided by magnification factor), which fulfills the requirement of Nyquist theorem in this regard. In Fig. 3(c), the horizontal resolution is worse than the vertical resolution. Such an effect is due to the high order aberration of the lens, such as coma. We also note that, due to the pixel-response differences of the linear CCD, line-artifact is present in the raw scanning data [13]. This effect can be eliminated by performing a simple normalization process: 1) capture a reference image without any sample; 2) normalized the raw scanning image of the sample with the reference image. In this process, the reference image is sample-independent, i.e., it can be used for any sample.

In the second experiment, we want to characterize the field curve of the imaging system. Our sample was a chrome mask (1.8 cm * 1.8 cm) with a hole-array on it (fabricated by lithography). The size of the hole was about 1 µm in diameter and periodicity of the hole-array was 30 µm. The light source of this experiment is the same as before (a green LED with a 20 nm spectrum bandwidth). First, we captured a series of images as we mechanically shifted the chrome mask into different z positions; then, we analyzed the spot size to locate the best focal plane for different FOVs, with the result shown in Fig. 4 (for example, at 50% FOV, the best focal plane locates at z= 6 µm). The displacement of the best focal plane is directly related to field curve of the imaging system. Remarkably, the result shows that the field curvature is relatively small (maximum observed of 12 micron z-displacement) over the entire FOV under the monochromatic illumination.

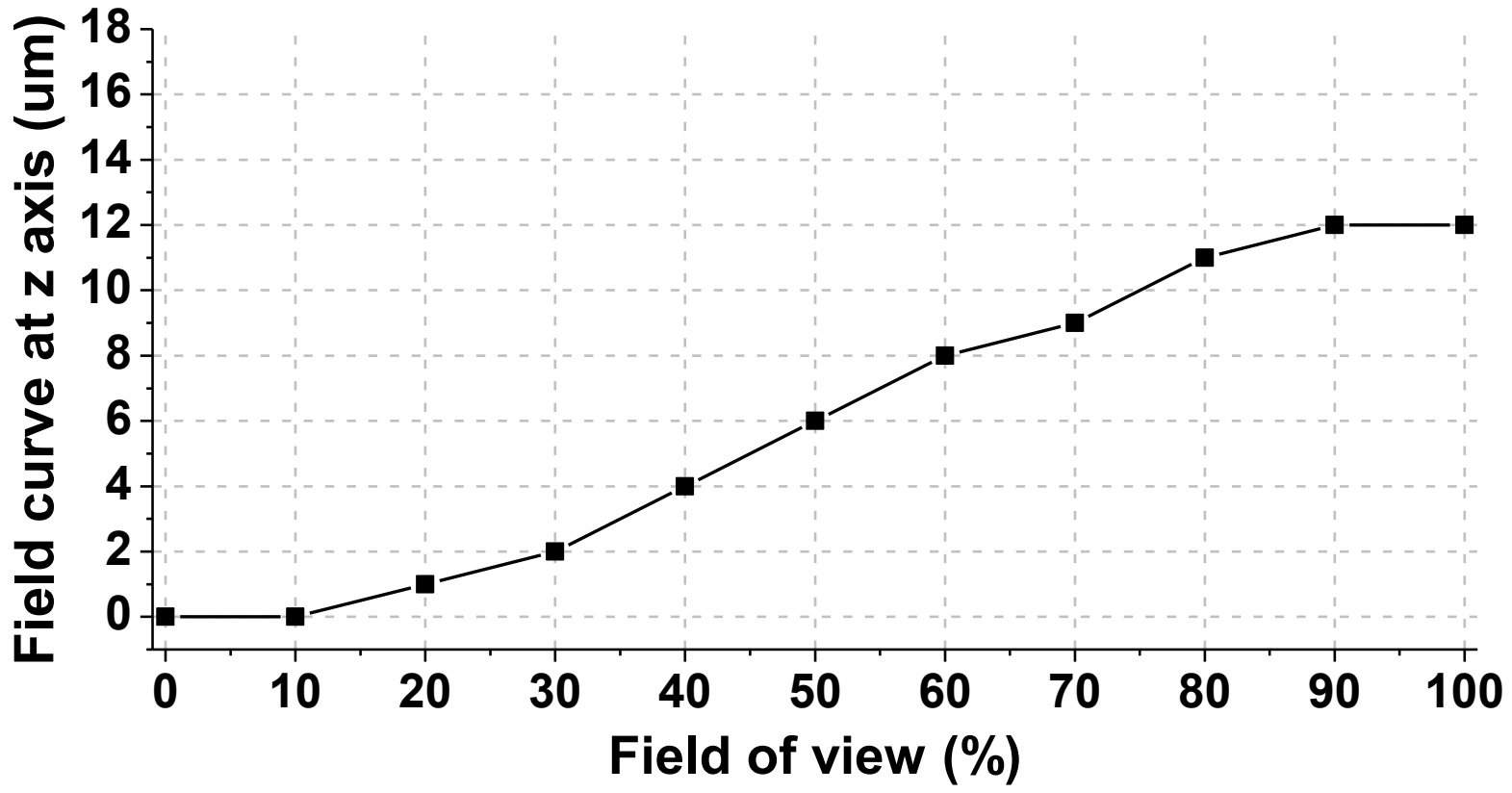

Fig. 4 Displacement of the best focal plane of different FOVs (from center to edge FOV). In this figure, 100% in x-axis corresponds to 10 mm.

## 5. Imaging of a blood smear and a pathology slide

We next used our system for imaging demonstration. First, we acquired a monochromatic image of a human blood smear using a green LED light source. The sample was automatically tuned to its in-focus position based on the automatic focusing scheme described in Section. 2. Fig. 5 shows the acquired image, where the scanner and magnification setting is the same as before. There are 200 times difference in the scale bars between Fig. 5(a) and (b3). In order to appreciate the wide FOV capability, we also prepare a video showing the entire and the magnified view of the image (Media 1).

The proposed platform can also be used for color imaging. We used R/G/B diffused LED light sources for three color illuminations, similar to Ref. [14]. The sample slide was automatically tuned into its in-focus position for each color illumination (Δz=3 µm for the blue LED and -12 µm for the red LED). Three images (for R/G/B) are separately acquired, normalized, aligned [15] and then combined into the final color image. Fig. 6 shows the acquired color image of a pathology slide (human metastatic carcinoma to liver, Carolina Biological Supply). Fig. 6(a) shows the wide FOV image of the pathology sample. Fig.6 (b1), (b2) and (c1), (c2) are the corresponding expanded view for Fig. 6(a).

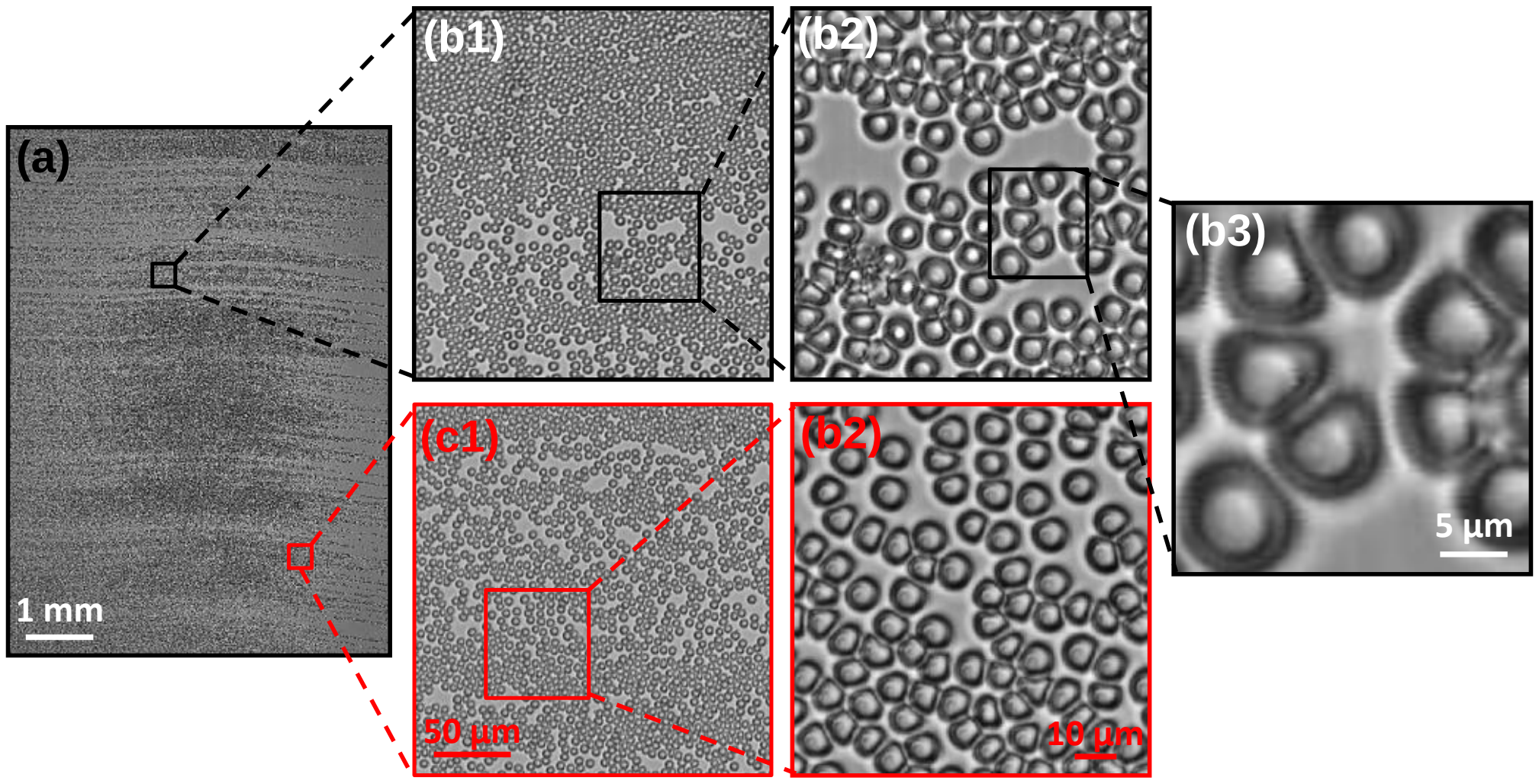

Fig. 5 (Media 1) 0.54 monochromatic gigapixel image of a blood smear. (a) The full frame of the captured image. (b1), (b2), (b3) and (c1), (c2) are the expanded view of the (a).

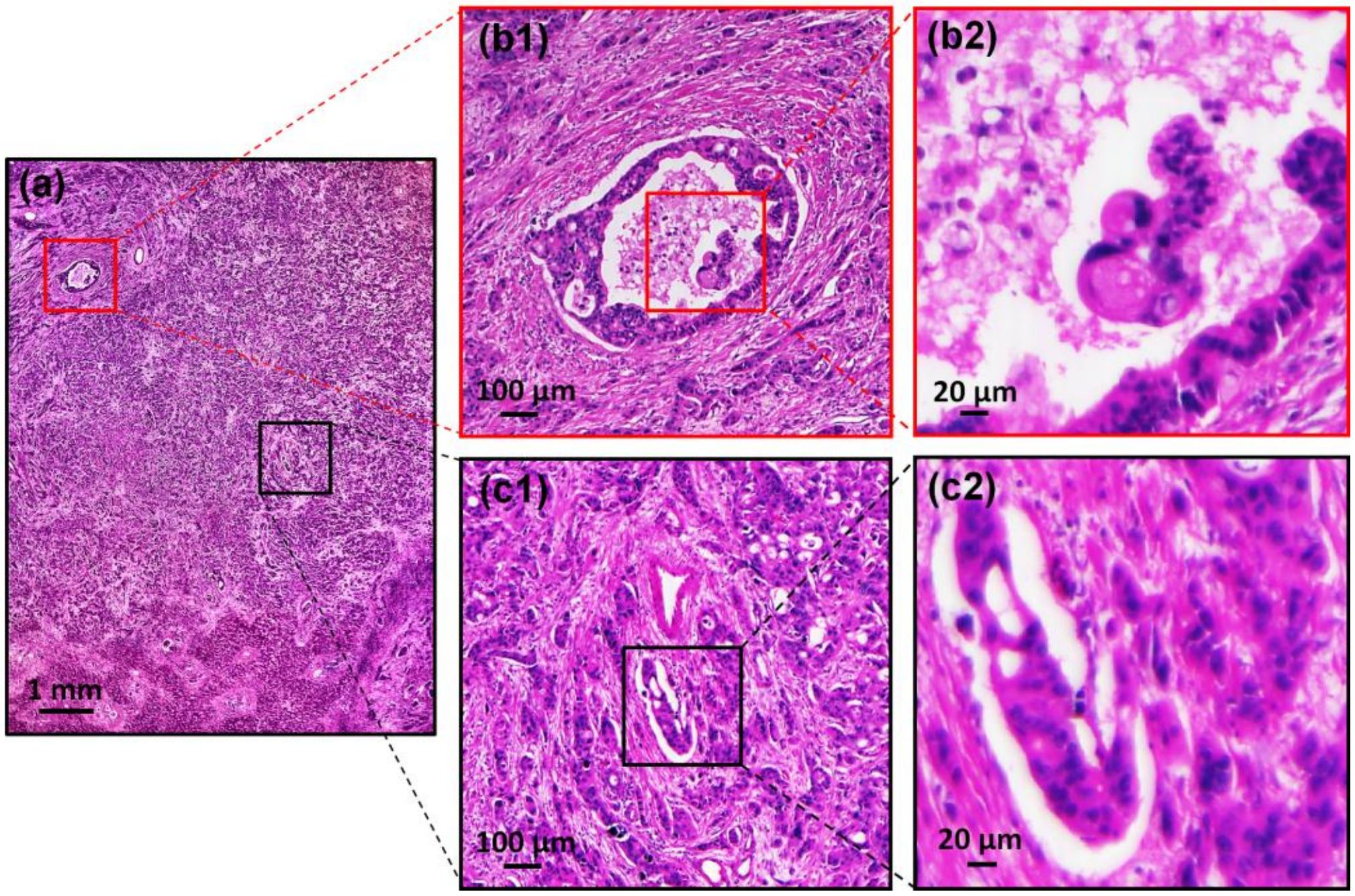

Fig. 6 0.54 gigapixel color image of a pathology slide (human metastatic carcinoma to liver). (a) The full frame of the acquired image. (b1), (b2) and (c1), (c2) are the expanded view of (a).

## 6. Conclusion

In summary, we report a wide-FOV (10 mm * 7.5 mm) microscopy system which can generate a 0.54 gigapixel microscopy image with 0.77 μm resolution across the entire FOV.

We note that there are other large-format professional camera lenses for even larger FOV (for example, 35 mm in diameter). The bottom line we want to convey in this paper is that lenses from the photography/industry community may provide a potential solution for high throughput microscopy imaging. Interested readers can choose their lenses based on the balance between the price and the performance.

It is interesting to contrast the SBP and resolution of our demonstrated system versus those of the conventional microscope. As shown in Fig. 7, the effective SBP of our system is more than one order of magnitude greater than those of the microscope objectives. Compared to a typical 10X and 4X objectives, our system has both superior SBP and resolution.

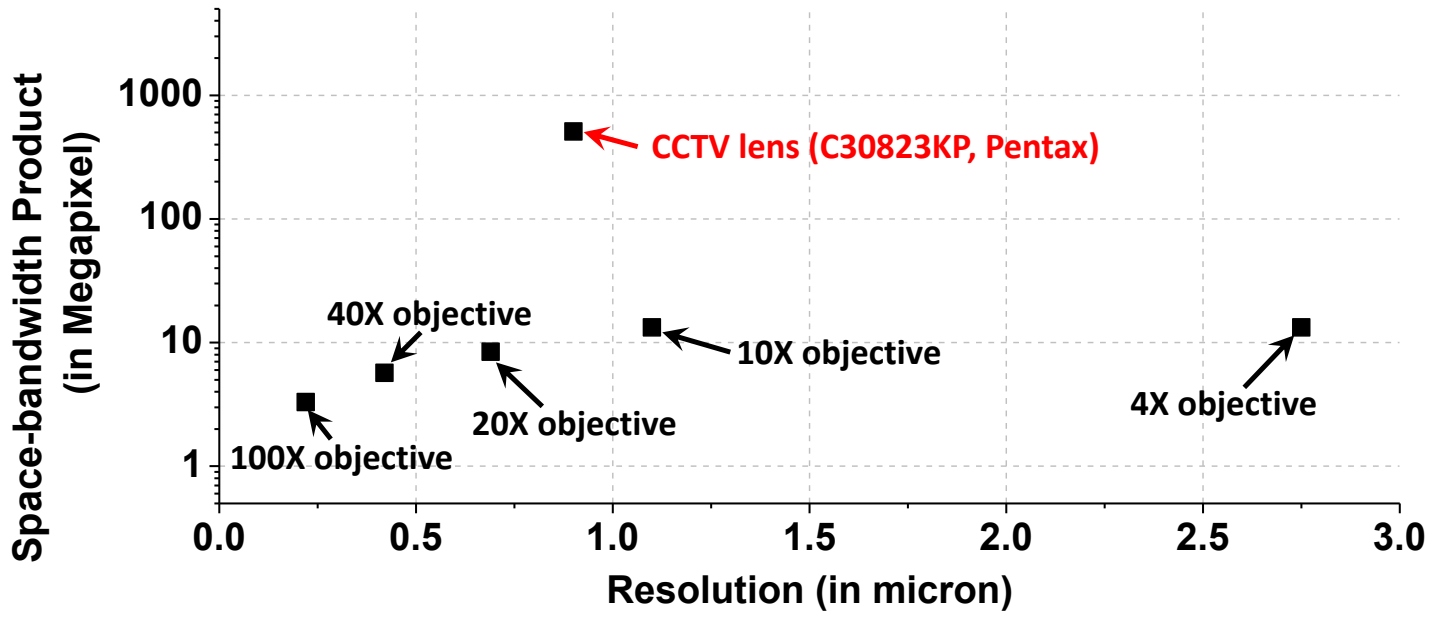

Fig. 7 The SBP-resolution summary for microscope objectives and the proposed system.

We note that one important limitation of the system is the scanning speed. A full-scan at 2400 dpi scanning resolution with USB 2.0 link took about 10 minutes. There are three strategies to address this issue: 1) use other high speed scanners; 2) use multiple scanners for parallelization (we can take out the linear CCDs and its housing components from multiple scanners and assemble them into one scanner); 3) use other scanning devices such as the digital scanning back (for example, a commercial available digital scanning back takes 29 seconds to capture a 0.312 gigapixel image [16]). Our future directions include: 1) improve the speed of setup using the strategies discussed above; 2) incorporate other imaging functionaries such as 3D, darkfield and phase imaging into the proposed wide FOV platform [17, 18].

**Acknowledgements**

We are grateful for the constructive discussions and generous help from Mr. Roarke Horstmeyer, Dr. Jigang Wu, and Mr. Zheng Li from Caltech. We acknowledge funding support from the United States Department of Defense under grant W81XWH-09-1-0051.